\documentclass[aps,prb,twocolumn,superscriptaddress,showpacs]{revtex4}

\usepackage{graphicx}           
\usepackage{acronym}                  
\usepackage{subfig}  
\usepackage{gensymb}                  
\usepackage{textcomp}                 
\usepackage{pinlabel}

\begin{document}

\title{(Ga,Mn)As based superlattices and the search for antiferromagnetic interlayer coupling}

\author{A.D. Giddings}
\affiliation{School of Physics and Astronomy, University of Nottingham, Nottingham NG7 2RD, UK}

\author{T. Jungwirth}
\affiliation{Institute of Physics ASCR, v.v.i., Cukrovarnick\'a 10, 162 53 Praha 6, Czech Republic}
\affiliation{School of Physics and Astronomy, University of Nottingham, Nottingham NG7 2RD, UK}

\author{B.L. Gallagher}
\affiliation{School of Physics and Astronomy, University of Nottingham, Nottingham NG7 2RD, UK}

\begin{abstract}
Antiferromagnetic interlayer coupling in dilute magnetic semiconductor superlattices could result in the realisation of large magnetoresistance effects analogous to the giant magnetoresistance seen in metallic multilayer structures. In this paper we use a mean-field theory of carrier induced ferromagnetism to explore the multidimensional parameter space available in (Ga,Mn)As based superlattice systems. Based on these investigations we examine the feasibility of creating a superlattice that exhibits antiferromagnetic coupling and suggest potentially viable recipes.
\end{abstract}


\maketitle

\section{Introduction}

The exciting new prospect of spin based electronics, known as spintronics, was initiated in 1988 with the discovery of \ac{GMR} in metallic multilayer structures.\cite{baibich_giant_1988,velu_enhanced_1988,binasch_enhanced_1989}
These structures consist of interposed \ac{FM} and non-\ac{FM} layers.
When  the magnetisation of adjacent \ac{FM} layers is aligned in antiparallel directions, enhanced spin scattering of carriers causes an increased electrical resistance through the layers, while when they are parallel the resistance is lower. Although typical \ac{GMR} devices today consist of a trilayer structure with a pinned magnetic layer and one in which the magnetisation is free to rotate, another method of implementation is with a superlattice structure where adjacent layers have an antiparallel magnetisation unless an external field is applied to align them.

In multilayer structures containing ferromagnetic layers, in addition to the ferromagnetic order within the layers, there can also exist magnetic exchange between the layers. The mechanism which causes the magnetic order between the layers is known as \ac{IEC}, and has been shown in metallic systems to be due to the spin polarisation of conduction carriers.\cite{bruno_theory_1995}

Because the \ac{IEC} energy considers the spin dependent changes in total energy, it thus determines which magnetic alignment of adjacent layers is energetically favourable. Although complicated helical arrangements can exist,\cite{nunez_helical_1995} typically the interlayer exchange coupling will either be \ac{FM}, where there is a parallel alignment of magnetisation, or \ac{AFM} where there is an antiparallel alignment. Therefore, in such a system, achieving \ac{AFM} interlayer coupling is of high importance for technological applications.

In addition to existing in metal systems, \ac{IEC} is a generic property of magnetic multilayers, and \ac{AFM} \ac{IEC} has even been demonstrated in non-metallic \ac{FM} semiconductor systems based on all-semiconductor EuS/PbS superlattices.\cite*{kepa_antiferromagnetic_2001}
\ac{AFM} \ac{IEC} in \ac{DMS} based superlattices was theoretically predicted in 1999 using a ${\bf k} \cdot {\bf p}$ kinetic-exchange model for carrier mediated ferromagnetism.\cite{jungwirth_interlayer_1999} This approach considers delocalised charge and adds extra modulation induced by spin-polarised effects. A large \ac{MR} was predicted due to the large difference in miniband dispersion for the cases of ferromagnetically and antiferromagnetically aligned layers. Recently, \ac{IEC} has been further explored using a tight-binding model.\cite{sankowski_interlayer_2005} This complementary microscopic approach, although not self-consistent, takes into account atomic orbitals for all the constituent atoms, leading to more accurate descriptions of the band structure.
Despite the different approaches used, both methods provide qualitatively similar results for the \ac{IEC} which shows oscillatory \ac{RKKY}-like behaviour.

Although \ac{IEC} has been shown to exist in \ac{DMS} systems based on (Ga,Mn)As/(Al,Ga)As trilayers,\cite{chiba_magnetoresistance_2000} there have been no reports of \ac{AFM} interlayer coupling. Experimental work into (Ga,Mn)As based multilayer and superlattice structures has only succeeded in demonstrating \ac{FM} \ac{IEC}.\cite{kepa_ferromagnetism_2001,chung_possible_2004} In order to test the prediction of a phenomenon analogous to \ac{GMR} in metals in \ac{DMS} materials, with a potentially much greater \ac{MR} ratio, it is essential that \ac{AFM} interlayer coupling is obtained.

The aim of this study is to provide a comprehensive description of the multidimensional parameter space available in these \ac{DMS} superlattice systems, in order to identify optimal parameters for realising an antiferromagnetically coupled system. Because the interlayer coupling is mediated by carriers, a ${\bf k} \cdot {\bf p}$ approach is more practical for exploring a wide range of parameter values. The limitation of this approach is that a single parabolic band approximation is used, sacrificing full quantitative accuracy for qualitative descriptions of a wide range of systems. Subtleties of the band-structure and spin-orbit effects are neglected. However, qualitative agreement with the data published in Ref. \onlinecite{sankowski_interlayer_2005} at least partially justifies this approach.

The organisation of this paper is as follows: first the details of the theoretical modelling of a \ac{DMS} based superlattice system and the numerics of the self-consistent mean field calculations will be shown.
Next, the results, which will primarily consider (Ga,Mn)As based superlattice systems with either GaAs or (Al,Ga)As non-magnetic spacer layers, will be presented.
Finally, in the discussion, suggestions for recipes for superlattice systems in which antiferromagnetic interlayer coupling may occur will be given.

\section{Theoretical modelling}

Our calculations are based on the Zener kinetic-exchange model\cite{zener_interaction_1951} description of magnetic interactions in Mn-doped III-V semiconductor structures. Microscopically, the kinetic-exchange between the local Mn moments and itinerant hole spins originates from the $p$-$d$ orbital hybridisation.\cite{jungwirth_theory_2006} This model provides a good description of ferromagnetism in bulk (Ga,Mn)As when the detailed structure of the valence band is taken into account.

An intuitive picture of the \ac{IEC} in (III,Mn)V/III-V multilayer structures can be obtained by the perturbative mapping of the kinetic-exchange model onto an effective interaction between local moments, following the \ac{RKKY} approach.\cite{jungwirth_interlayer_1999} The \ac{RKKY} theory can be expected to provide useful predictions for structures close to a model pseudo-1D system consisting of alternating thin ferromagnetic layers and non-magnetic spacer layers such that there is small coupling and low carrier polarisation.\cite{seitz_solid_1968} The \ac{RKKY} range function falls off asymptotically with $\frac{\sin(2 k_F d)}{d^2}$, where $k_F$ is the carrier wave vector and $d$ is the distance between the magnetic layers. Thus,
the \ac{RKKY} theory shows that the coupling can have an oscillatory form.

The Zener kinetic-exchange model for homogeneous (Ga,Mn)As was generalized in Ref. \onlinecite{jungwirth_interlayer_1999} in order to account for the RKKY-like oscillatory effects in the inter-(Ga,Mn)As coupling in (Ga,Mn)As based ferromagnetic/non-magnetic superlattices on a more quantitative level.
In this model the band structure is solved using the kinetic-exchange model and a parabolic band ${\bf k} \cdot {\bf p}$ effective mass approximation. In the Hamiltonian the magnetic moments are accounted for through the $p$-$d$ kinetic-exchange interaction between Mn spins and hole spins which is parametrized by a constant $J_{pd}$ and treated in the mean-field virtual crystal approximation. The value of $J_{pd}$ can be experimentally determined, and modern estimates of this value place it at $55$ meV$\,$nm$^3$.\cite{sinova_magneto-transport_2004} To account for the inhomogeneity, a standard formalisation of the \ac{LSDA} using the Kohn-Sham equations for inhomogeneous systems is used in the band structure calculations.\cite{vosko_accurate_1980}
Hole mass is $m^* = 0.5 m_e$ and the spin of local Mn moments is $S = \frac{5}{2}$ at $T = 0$ K.
Thermodynamics are treated on a mean field level.

In order to find the normalized wavefunction for a given energy, Bloch's theorem is used to solve the one-dimensional time-independent spin-dependent Schr\"{o}dinger equation:

\begin{equation}
(\frac{p^2}{2m^*} + V_\sigma(z)) \Psi_{k,n,\sigma} (z) = E_{k,n,\sigma} \Psi_{k,n,\sigma} (z) ,
\end{equation}

which we shall rewrite as


\begin{eqnarray}
\nonumber \frac{d^2 \Psi_{k,n,\sigma}}{dz^2} &=& \frac{2m^*}{\hbar^2}(V_\sigma(z)-E_{k,n,\sigma}) \Psi_{k,n,\sigma} (z) , \\ 
\Psi''(z) &=& f \Psi(z) ,
\label{eq:schro}
\end{eqnarray}

where $k$ is the wavevector, $n$ is the subband index, and $\sigma$ is the spin index and $f = \frac{2m^*}{\hbar^2} (V_\sigma(z) - E)$.

The Bloch function

\begin{equation}
\Psi_{k,n,\sigma} (z) = u_{k,n,\sigma} (z) e^{(ikz)} ,
\label{eq:bloch}
\end{equation}

gives the solutions of the Schr\"{o}dinger equation for a periodic potential.

For this system, the explicit form of the Hamiltonian for the spin-dependant potential $V_\sigma(z)$ is given by

\begin{equation}
V_\sigma(z) = V_H + V_{xc,\sigma} + V_b - \frac{\sigma}{2}[g^* \mu_B B + h_{pd}(z)] ,
\end{equation}

where $V_H$ is the Hartree (electrostatic) potential, given by the Poisson equation, $V_{xc,\sigma}$ is the spin-dependent exchange-correlation potential given by the \ac{LSDA} equation, $V_b$ is the band-offset, $g^*$ is the free-carrier g-factor and $h_{pd}$ is the mean-field kinetic-exchange interaction.\cite{jungwirth_interlayer_1999} The \ac{IEC} energy, $E_c$, is defined as the difference in energy between the \ac{FM} and \ac{AFM} states per superlattice period.

Let us suppose that our one-dimensional lattice has a period $d_{n+m}$ and consider now the solution only at $N$ evenly distributed discrete points on the $z$-axis with a separation $h$. The wavefunction at each point is denoted as $\Psi(z)$.
By Taylor's theorem, the second-order approximations for $\Psi(z+h)$ and $\Psi(z-h)$ are

\begin{eqnarray}
\label{eq:taylor1}
\Psi(z+h) &=& \Psi(z) + h \Psi'(z) + \frac{h^2}{2} \Psi''(z) ; \\
\Psi(z-h) &=& \Psi(z) - h \Psi'(z) + \frac{h^2}{2} \Psi''(z).
\label{eq:taylor2}
\end{eqnarray}

Taking the sum of Eqs.~(\ref{eq:taylor1}) and (\ref{eq:taylor2}) we obtain 

\begin{equation}
h^2 \Psi''(z) = \Psi(z+h) + \Psi(z-h) - 2\Psi(z).
\label{eq:taylor3}
\end{equation}



Substituting the Schr\"{o}dinger equation from Eq.~(\ref{eq:schro}) into Eq.~(\ref{eq:taylor3}) and rearranging gives the wavefunction at a given point as a linear combination of the wavefunctions at the two previous points:

\begin{equation}
\Psi(z+h) = (h^2 f + 2)\Psi(z) - \Psi(z-h).
\end{equation}

This linear transformation can be represented as a transfer matrix, ${\bf M}_n$,
such that

\begin{eqnarray}
{\bf M}_n
\left( \begin{array}{ccc}
\Psi_n \\ \Psi_{n-1} \\
\end{array} \right)
= 
\left( \begin{array}{ccc}
m_{n,11} & m_{n,12} \\ m_{n,21} & m_{n,22} \\
\end{array} \right)
\left( \begin{array}{ccc}
\Psi_n \\ \Psi_{n-1} \\
\end{array} \right) \nonumber\\
=
\left( \begin{array}{ccc}
\Psi_{n+1} \\ \Psi_n \\
\end{array} \right),
\end{eqnarray}

where $\Psi_n$ is the wavefunction at the $n^{th}$ $z$-point. By inspection we see that $m_{n,11} = h^2 f + 2$, $m_{n,12} = -1$, $m_{n,21} = 1$ and $m_{n,22} = 0$. It is worth noting here that the determinant of each ${\bf M}_n$, $\det({\bf M}_n) = 1$. The product of the $N$ transfer matrices $\prod_{n=1}^N {\bf M}_n = {\bf M}^T$ represents the transformation from $\Psi_0$ to $\Psi_N$.
By Bloch theorem's periodic boundary condition, Eq.~(\ref{eq:bloch}), this transformation can be written

\begin{eqnarray}
{\bf M}^T
\left( \begin{array}{ccc}
\Psi_1 \\ \Psi_0 \\
\end{array} \right)
=
\left( \begin{array}{ccc}
m^T_{11} & m^T_{12} \\ m^T_{21} & m^T_{22} \\
\end{array} \right)
\left( \begin{array}{ccc}
\Psi_1 \\ \Psi_0 \\
\end{array} \right)
=
\left( \begin{array}{ccc}
\Psi_{N+1} \\ \Psi_N \\
\end{array} \right) \nonumber \\
= e^{ika}
\left( \begin{array}{ccc}
\Psi_1 \\ \Psi_0 \\
\end{array} \right).
\end{eqnarray}

Therefore,

\begin{eqnarray}
0 & = & \det ({\bf M}^T - e^{ika} {\bf I}_2)  \nonumber \\
& = & (m^T_{11} - e^{ika})(m^T_{22} - e^{ika}) - m^T_{12} m^T_{21}\nonumber \\
& = & m^T_{11} m^T_{22} - m^T_{12} m^T_{21} - e^{ika}(m^T_{11} + m^T_{22}) + e^{2ika} \nonumber \\
& = & \det ({\bf M}^T) -  e^{ika} \textnormal{Tr} ({\bf M}^T) +  e^{2ika}.
\label{eq:determinant}
\end{eqnarray}

Since the determinant of ${\bf M}_n$ is 1, then the determinant of any product of ${\bf M}_n$, for any $n$, will also have a determinant of 1, hence $\det({\bf M}^T)=1$. Substituting this into Eq.~(\ref{eq:determinant}) gives

\begin{equation}
1- e^{ika} \textnormal{Tr}({\bf M}^T) +  e^{2ika} =0 \nonumber \\
\end{equation}
\begin{eqnarray}
\textnormal{Tr}({\bf M}^T) &=& e^{ika} + e^{-ika} \nonumber \\
 &=& 2 \cos(ka)
\end{eqnarray}

For a given energy, wavevector $k$ can thus be found by

\begin{equation}
k= \frac{1}{a} \arccos (\frac{1}{2} \textnormal{Tr}({\bf M}^T)) ,
\end{equation}

and the corresponding wavefunction can be found similarly.

\section{Results}

In the \ac{RKKY} model of interlayer exchange the oscillations occur as a function of $k_F d$, where $k_F$ is the Fermi wave vector and $d$ is the separation between the two-dimensional magnetic planes.\cite{yafet_ruderman-kittel-kasuya-yosida_1987}
In our model we shall denote $d_n$ as the width of the non-magnetic layers, corresponding to $d$ from the \ac{RKKY} model, and $d_m$ as the width of the magnetic layers. The length of a GaAs unit cell is labelled $a_0$ and has a value of 0.565 nm. We shall also define the average Fermi wave vector $\bar{k}_F$ as

\begin{equation}
\bar{k}_F = (3 \pi^2 \bar{N}_{3D})^{\frac{1}{3}} ,
\end{equation}

corresponding to the Fermi vector $k_F$ in the ideal \ac{RKKY} model with a parabolic band. The average 3D carrier concentration $\bar{N}_{3D}$ is defined as 

\begin{equation}
\bar{N}_{3D} = \frac{1}{d_{n+m}} \int_{\textnormal{\scriptsize unit cell}} N_{3D}(z) \,dz = \frac{N_{2D}}{d_{n+m}} .
\end{equation}

\begin{figure*}
  \centering
  \labellist%
    \footnotesize%
    \pinlabel $0$ [B] at 84, 63
    \pinlabel $2$ [B] at 98, 63
    \pinlabel $4$ [B] at 113, 63
    \pinlabel $6$ [B] at 127, 63
    \pinlabel $8$ [B] at 142, 63
    \pinlabel $10$ [B] at 157, 63
    \pinlabel $12$ [B] at 171, 63
    \pinlabel $2$ [B] at 200, 63
    \pinlabel $4$ [B] at 215, 63
    \pinlabel $6$ [B] at 229, 63
    \pinlabel $8$ [B] at 244, 63
    \pinlabel $10$ [B] at 258, 63
    \pinlabel $12$ [B] at 273, 63
    \pinlabel $2$ [B] at 302, 63
    \pinlabel $4$ [B] at 317, 63
    \pinlabel $6$ [B] at 331, 63
    \pinlabel $8$ [B] at 346, 63
    \pinlabel $10$ [B] at 360, 63
    \pinlabel $12$ [B] at 375, 63
    \pinlabel $14$ [B] at 389, 63
    \pinlabel {$d_n$ / $\frac{1}{2}a_0$} [B] at 135, 50
    \pinlabel {$d_n$ / $\frac{1}{2}a_0$} [B] at 237, 50
    \pinlabel {$d_n$ / $\frac{1}{2}a_0$} [B] at 338, 50
    \pinlabel $-300$ [r] at 84, 75
    \pinlabel $-200$ [r] at 84, 109
    \pinlabel $-100$ [r] at 84, 143
    \pinlabel $0$ [r] at 84, 177
    \pinlabel $0.2$ [r] at 84, 192
    \pinlabel $0.4$ [r] at 84, 208
    \pinlabel $0.6$ [r] at 84, 223
    \pinlabel $0.8$ [r] at 84, 239
    \pinlabel $1.0$ [r] at 84, 255
    \pinlabel $1.2$ [r] at 84, 270
    \pinlabel \rotatebox{90}{$V_\sigma$ / meV} at 49, 126
    \pinlabel \rotatebox{90}{$N_{3D} (z)$ / $10^{20}$ cm$^{-3}$} at 49, 224
    \hair 8pt
    \pinlabel (a) [tl] at 84, 278
    \pinlabel (b) [tl] at 186, 278
    \pinlabel (c) [tl] at 287, 278
    \hair 4pt
    \pinlabel {up spin} [r] at 157, 107
    \pinlabel {down spin} [r] at 157, 95
    \pinlabel {total} [r] at 157, 83
    \pinlabel {\shortstack{magnetic \\ layer}} at 315 90
    \pinlabel {\shortstack{non-magnetic \\ layer}} at 340 146
  \endlabellist%
  \includegraphics[width=0.76\textwidth]{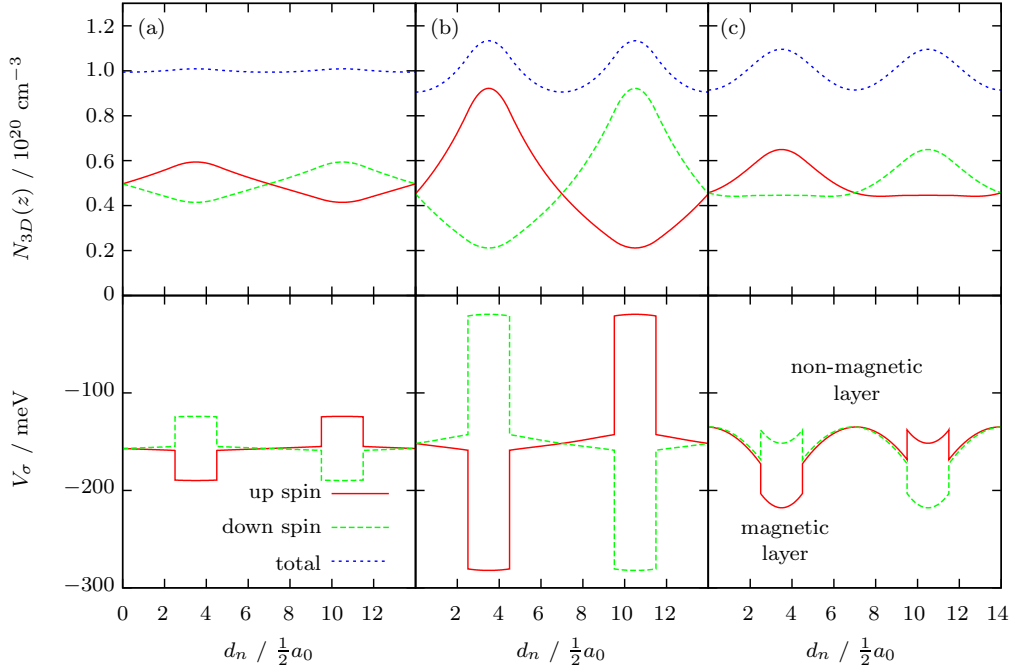}
  \caption{(colour online) The self-consistent charge distribution, $N_{3D}$, and potentials, $V_\sigma$, for a double unit cell of three different (Ga,Mn)As/GaAs based superlattice structures in an \ac{AFM} state. $d_m / \frac{1}{2}a_0 = 2$, $d_n / \frac{1}{2}a_0 = 5$ and $\bar{N}_{3D} = 10^{20}$ cm$^{-3}$ in each case. $V_\sigma = 0$ eV corresponds to the Fermi level.
(a) 2\% Mn doping and a uniform impurity concentration, (b) 8\% Mn doping and a uniform impurity concentration and (c) 2\% Mn doping but no impurities in the non-magnetic layer.}
  \label{fig:GaAs_potentials}
\end{figure*}

\begin{figure*}
  \centering
  \labellist%
    \footnotesize%
    \pinlabel $0$ [B] at 84, 63
    \pinlabel $2$ [B] at 98, 63
    \pinlabel $4$ [B] at 113, 63
    \pinlabel $6$ [B] at 127, 63
    \pinlabel $8$ [B] at 142, 63
    \pinlabel $10$ [B] at 157, 63
    \pinlabel $12$ [B] at 171, 63
    \pinlabel $2$ [B] at 200, 63
    \pinlabel $4$ [B] at 215, 63
    \pinlabel $6$ [B] at 229, 63
    \pinlabel $8$ [B] at 244, 63
    \pinlabel $10$ [B] at 258, 63
    \pinlabel $12$ [B] at 273, 63
    \pinlabel $4$ [B] at 303, 63
    \pinlabel $8$ [B] at 319, 63
    \pinlabel $12$ [B] at 334, 63
    \pinlabel $16$ [B] at 350, 63
    \pinlabel $20$ [B] at 366, 63
    \pinlabel $24$ [B] at 381, 63
    \pinlabel {$d_n$ / $\frac{1}{2}a_0$} [B] at 135, 50
    \pinlabel {$d_n$ / $\frac{1}{2}a_0$} [B] at 237, 50
    \pinlabel {$d_n$ / $\frac{1}{2}a_0$} [B] at 338, 50
    \pinlabel $-500$ [r] at 84, 75
    \pinlabel $-400$ [r] at 84, 95
    \pinlabel $-300$ [r] at 84, 116
    \pinlabel $-200$ [r] at 84, 136
    \pinlabel $-100$ [r] at 84, 156
    \pinlabel $0$ [r] at 84, 177
    \pinlabel $0.4$ [r] at 84, 198
    \pinlabel $0.8$ [r] at 84, 219
    \pinlabel $1.2$ [r] at 84, 241
    \pinlabel $1.6$ [r] at 84, 262
    \pinlabel \rotatebox{90}{$V_\sigma$ / meV} at 49, 126
    \pinlabel \rotatebox{90}{$N_{3D} (z)$ / $10^{20}$ cm$^{-3}$} at 49, 224
    \hair 8pt
    \pinlabel (a) [tl] at 84, 278
    \pinlabel (b) [tl] at 186, 278
    \pinlabel (c) [tl] at 287, 278
    \hair 4pt
    \pinlabel {up spin} [r] at 157, 107
    \pinlabel {down spin} [r] at 157, 95
    \pinlabel {total} [r] at 157, 83
  \endlabellist%
  \includegraphics[width=0.76\textwidth]{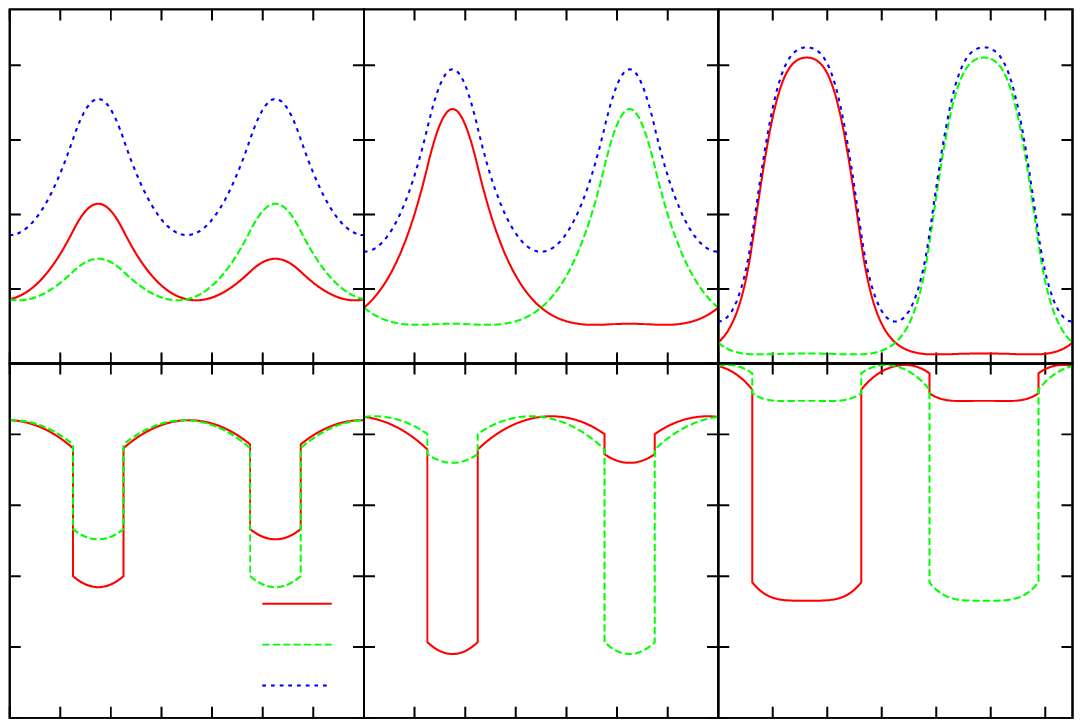}
  \caption{(colour online) The self-consistent charge distribution, $N_{3D}$, and potentials, $V_\sigma$, for a double unit cell of three different (Ga,Mn)As/(Al,Ga)As based superlattice structures in an \ac{AFM} state. $d_n / \frac{1}{2}a_0 = 5$, $\bar{N}_{3D} = 10^{20}$ cm$^{-3}$ and the Al concentration is 30\% in each case. $V_\sigma = 0$ eV corresponds to the Fermi level.
(a) $d_m / \frac{1}{2}a_0 = 2$ and 2\% Mn doping (b) $d_m / \frac{1}{2}a_0 = 2$ and 8\% Mn doping (c) $d_m / \frac{1}{2}a_0 = 8$ and 8\% Mn doping.} 
  \label{fig:AlGaAs_potentials}
\end{figure*} 

The superlattice structures being considered in this paper consist of thin (Ga,Mn)As layers interposed with non-magnetic spacer layers. The primary structural parameters that can be changed are the widths of the layers and their composition. Fig.~\ref{fig:GaAs_potentials}(a) shows the calculated self-consistent charge distribution and potentials for a simple case with a low moment concentration (2\%), and the spacer layers are thicker than the magnetic layers. There is a uniform impurity concentration of acceptors throughout the structure, either magnetic, like Mn in the magnetic layers, or from non-magnetic dopants in the spacer layers. The polarisation of carriers is low in this case, and they have an almost totally uniform distribution. These figures show cases where the number of monolayers of non-magnetic layer $d_n / \frac{1}{2} a_0 = 5$ and average 3D carrier concentration $\bar{N}_{3D} = 10^{20}$ cm$^{-3}$. The Fermi energy is at $V_\sigma = 0$ eV. The interlayer coupling is in an assumed \ac{AFM} state, although this may not be the energetically favoured state for such a system. 

The effect of increasing the number of moments in the magnetic layers is shown in Fig.~\ref{fig:GaAs_potentials}(b), where the Mn doping is 8\%. As expected, the increased moment concentration increases the size of spin splitting in the magnetic layers, which increases the carrier polarisation. Additionally, the potential due to the magnetic ordering causes a redistribution of carriers to occur, increasing the concentration in the magnetic layer. Another way to cause carrier redistribution is to remove the doping from the non-magnetic layer, which is shown in Fig.~\ref{fig:GaAs_potentials}(c), where the Mn concentration in the magnetic layers is again 2\%. Without a neutralising background charge, Coulomb repulsion opposes carrier redistribution into the non-magnetic layers. The resulting carrier distribution is similar to that of Fig.~\ref{fig:GaAs_potentials}(b). Although there is a greater concentration of carriers in the magnetic layers, the polarisation is not significantly increased over that seen in Fig.~\ref{fig:GaAs_potentials}(a) where there is a uniform charge concentration.

To cause stronger confinement of carrier to the magnetic layers, the non-magnetic layers can be made from (Al,Ga)As. The effects of this are shown in  Fig.~\ref{fig:AlGaAs_potentials}(a), where the carrier concentration in the centre of the magnetic layers is about double that of the centre of the non-magnetic layers. Although there is a slight increase in the polarisation of carriers over the previous two cases with 2\% Mn moment, this is increased significantly when more moments (8\%) are put in the magnetic layers, as shown in  Fig.~\ref{fig:AlGaAs_potentials}(b). Because of the strong confinement of carriers to the highly magnetic layers there is a very high polarisation of carriers; this effect is enhanced over that of the 8\% case with a doped GaAs spacer which was shown in Fig.~\ref{fig:GaAs_potentials}(b). Finally, Fig.~\ref{fig:AlGaAs_potentials}(c) shows a case where the non-magnetic layers are thinner than the magnetic layers, and also have a high Mn doping. Following the established trend, carriers are strongly polarised and tightly confined to the magnetic layers. In fact, with an almost total polarisation and a very high depletion of the non-magnetic layers, this structure represents an almost opposite case to that of Fig.~\ref{fig:GaAs_potentials}(a).

Bearing in mind these examples of how changing the structural properties can alter the electronic configuration of the superlattices, the effects of these changes on the \ac{IEC} will now be explored. This will be done in two parts. Firstly, GaAs based spacers, similar to those shown in Fig.~\ref{fig:GaAs_potentials} will be considered. The more extreme cases presented in Fig.~\ref{fig:AlGaAs_potentials} will be considered in the second half.

\subsection{GaAs spacer}

\begin{figure}
  \centering
  \labellist%
    \footnotesize%
    \pinlabel -10 [r] at 79 126
    \pinlabel 0 [r] at 79 141
    \pinlabel 10 [r] at 79 156
    \pinlabel 20 [r] at 79 171
    \pinlabel 30 [r] at 79 186
    \pinlabel 2 [tr] at 81 121
    \pinlabel 4 [tr] at 94 108
    \pinlabel 6 [tr] at 107 95
    \pinlabel 8 [tr] at 120 82
    \pinlabel 10 [tr] at 133 69
    \pinlabel $10^{19}$ [tl] at 133 69
    \pinlabel $10^{20}$ [tl] at 178 84
    \pinlabel $10^{21}$ [tl] at 223 99
    \pinlabel {$d_n$ / $\frac{1}{2}a_0$} [tr] at 97 85
    \pinlabel {$\bar{N}_{3D}$ / cm$^{-3}$} [tl] at 193 69
    \pinlabel \rotatebox{90}{$E_c$ / \micro J\,m$^{-2}$} at 52 156
  \endlabellist%
  \subfloat[$d_m / \frac{1}{2}a_0 = 2$, 2\% Mn concentration]%
    {\includegraphics[width=0.48\textwidth]{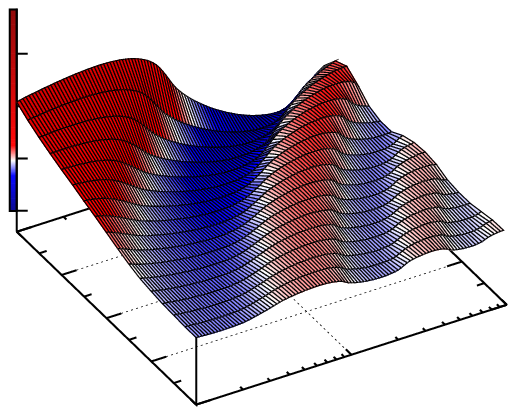}}\\
    \labellist%
    \footnotesize%
    \pinlabel -10 [r] at 79 126
    \pinlabel 0 [r] at 79 141
    \pinlabel 10 [r] at 79 156
    \pinlabel 20 [r] at 79 171
    \pinlabel 30 [r] at 79 186
    \pinlabel 2 [tr] at 81 121
    \pinlabel 4 [tr] at 94 108
    \pinlabel 6 [tr] at 107 95
    \pinlabel 8 [tr] at 120 82
    \pinlabel 10 [tr] at 133 69
    \pinlabel $10^{19}$ [tl] at 133 69
    \pinlabel $10^{20}$ [tl] at 178 84
    \pinlabel $10^{21}$ [tl] at 223 99
    \pinlabel {$d_n$ / $\frac{1}{2}a_0$} [tr] at 97 85
    \pinlabel {$\bar{N}_{3D}$ / cm$^{-3}$} [tl] at 193 69
    \pinlabel \rotatebox{90}{$E_c$ / \micro J\,m$^{-2}$} at 52 156
  \endlabellist%
  \subfloat[$d_m / \frac{1}{2}a_0 = 8$, 2\% Mn concentration]%
    {\includegraphics[width=0.48\textwidth]{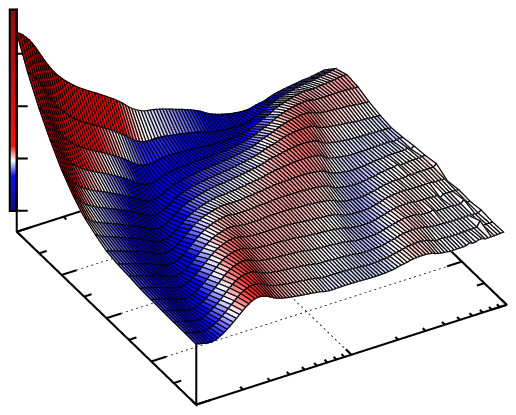}}%
  \caption{(colour online) The \acs{IEC} energy, $E_c$, of two (Ga,Mn)As/GaAs based superlattices as a function of the average 3D carrier concentration, $\bar{N}_{3D}$, and the number of monolayers of non-magnetic layer, $2 d_n / \frac{1}{2}a_0$. Positive (red coloured) values of $E_c$ indicate \ac{FM} interlayer coupling is energetically favourable and negative (blue coloured) values indicate \ac{AFM} is favourable. Both superlattices have a 2\% Mn doping in the magnetic layer and there is a uniform impurity concentration of acceptors through the structure.}%
  \label{fig:Doped_spacer_2}%
\end{figure}

\begin{figure}
  \centering%
    \labellist%
    \footnotesize%
    \pinlabel -100 [r] at 79 126
    \pinlabel 0 [r] at 79 141
    \pinlabel 100 [r] at 79 156
    \pinlabel 200 [r] at 79 171
    \pinlabel 300 [r] at 79 186
    \pinlabel 2 [tr] at 81 121
    \pinlabel 4 [tr] at 94 108
    \pinlabel 6 [tr] at 107 95
    \pinlabel 8 [tr] at 120 82
    \pinlabel 10 [tr] at 133 69
    \pinlabel $10^{19}$ [tl] at 133 69
    \pinlabel $10^{20}$ [tl] at 178 84
    \pinlabel $10^{21}$ [tl] at 223 99
    \pinlabel {$d_n$ / $\frac{1}{2}a_0$} [tr] at 97 85
    \pinlabel {$\bar{N}_{3D}$ / cm$^{-3}$} [tl] at 193 69
    \pinlabel \rotatebox{90}{$E_c$ / \micro J\,m$^{-2}$} at 52 156
  \endlabellist%
  \subfloat[$d_m / \frac{1}{2}a_0 = 2$, 8\% Mn concentration]%
    {\includegraphics[width=0.48\textwidth]{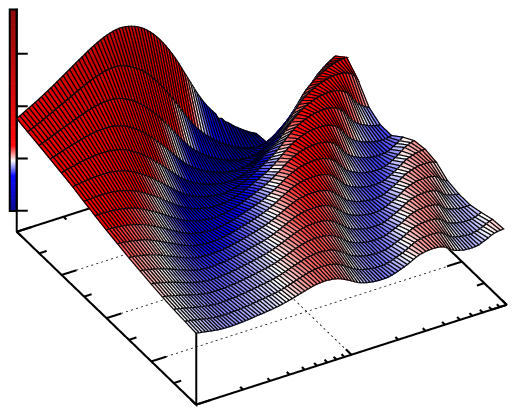}}\\
    \labellist%
    \footnotesize%
    \pinlabel -100 [r] at 79 126
    \pinlabel 0 [r] at 79 141
    \pinlabel 100 [r] at 79 156
    \pinlabel 200 [r] at 79 171
    \pinlabel 300 [r] at 79 186
    \pinlabel 2 [tr] at 81 121
    \pinlabel 4 [tr] at 94 108
    \pinlabel 6 [tr] at 107 95
    \pinlabel 8 [tr] at 120 82
    \pinlabel 10 [tr] at 133 69
    \pinlabel $10^{19}$ [tl] at 133 69
    \pinlabel $10^{20}$ [tl] at 178 84
    \pinlabel $10^{21}$ [tl] at 223 99
    \pinlabel {$d_n$ / $\frac{1}{2}a_0$} [tr] at 97 85
    \pinlabel {$\bar{N}_{3D}$ / cm$^{-3}$} [tl] at 193 69
    \pinlabel \rotatebox{90}{$E_c$ / \micro J\,m$^{-2}$} at 52 156
  \endlabellist%
  \subfloat[$d_m / \frac{1}{2}a_0 = 8$, 8\% Mn concentration]%
    {\includegraphics[width=0.48\textwidth]{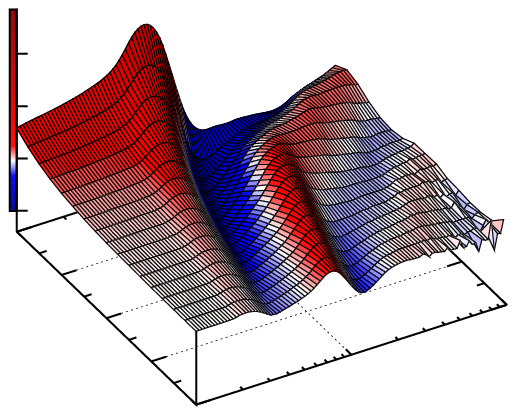}}%
  \caption{(colour online) The \acs{IEC} energy, $E_c$, of two (Ga,Mn)As/GaAs based superlattices as a function of the average 3D carrier concentration, $\bar{N}_{3D}$, and the number of monolayers of non-magnetic layer, $2 d_n / \frac{1}{2}a_0$. Both superlattices have an 8\% Mn doping in the magnetic layer and there is a uniform impurity concentration of acceptors through the structure.}%
  \label{fig:Doped_spacer_8}%
\end{figure}

First to be considered is a superlattice structure close to the \ac{RKKY} limit of infinitely thin magnetic layers surrounded by free unpolarised carriers. For this we shall use thin magnetic layers and a low magnetic moment concentration, as per Fig.~\ref{fig:GaAs_potentials}(a). In Fig.~\ref{fig:Doped_spacer_2}(a) the \ac{IEC} energy, $E_c$, is plotted against the 3D carrier concentration, $\bar{N}_{3D}$, and number of monolayers of GaAs in the non-magnetic spacer, $2 d_n / \frac{1}{2}a_0$. The magnetic (Ga,Mn)As layer is 2 monolayers thick and contains 2\% Mn local moment doping. There is a uniform acceptor density throughout the structure which gives an average hole concentration of $4.43 \times 10^{20}$ cm$^{-3}$. In this case there are oscillations as a function of both parameters, analogous to the $k_F d$ oscillations in the ideal quasi one-dimensional \ac{RKKY} model.
For the calculated IEC energy, $E_c$, positive values correspond to \ac{FM} interlayer coupling being energetically favourable, and negative values correspond to \ac{AFM} interlayer coupling being the favoured configuration.

The \ac{RKKY} like behaviour observed in Fig.~\ref{fig:Doped_spacer_2}(a) is consistent with the results obtained in the tight-binding approach\cite{sankowski_interlayer_2005} when the exchange coupling energy, $E_c$, is plotted against the two-dimensional carrier concentration, $N_{2D}$, for fixed layer thicknesses. However, it is worth noting that when the exchange coupling is plotted as a function of the non-magnetic spacer thickness, $d_n$, for a fixed $N_{2D}$ there are no apparent \ac{RKKY} oscillations. Because $N_{3D}$, and therefore $k_F$, is a function of $d_n$ these two parameters are not independent when $N_{2D}$ is fixed. This results in the oscillatory behaviour appearing to be suppressed.

\begin{figure}
  \centering
  \labellist%
    \footnotesize%
    \pinlabel $0$ [B] at 81, 60
    \pinlabel $2$ [B] at 105, 60
    \pinlabel $4$ [B] at 129, 60
    \pinlabel $6$ [B] at 153, 60
    \pinlabel $8$ [B] at 177, 60
    \pinlabel $10$ [B] at 201, 60
    \pinlabel $12$ [B] at 225, 60
    \pinlabel $14$ [B] at 249, 60
    \pinlabel $16$ [B] at 273, 60
    \pinlabel $18$ [B] at 297, 60
    \pinlabel $20$ [B] at 321, 60
    \pinlabel {$2 \bar{k}_F d_{n + 1}$} [B] at 201, 47 
    \pinlabel $-10$ [r] at 81, 72
    \pinlabel $-5$ [r] at 81, 106
    \pinlabel $0$ [r] at 81, 140
    \pinlabel $5$ [r] at 81, 174
    \pinlabel $10$ [r] at 81, 208
    \pinlabel $15$ [r] at 81, 242
    \pinlabel \rotatebox{90}{$E_c$ / \micro J\,m$^{-2}$} at 56, 157
  \endlabellist%
  \includegraphics[width=0.48\textwidth]{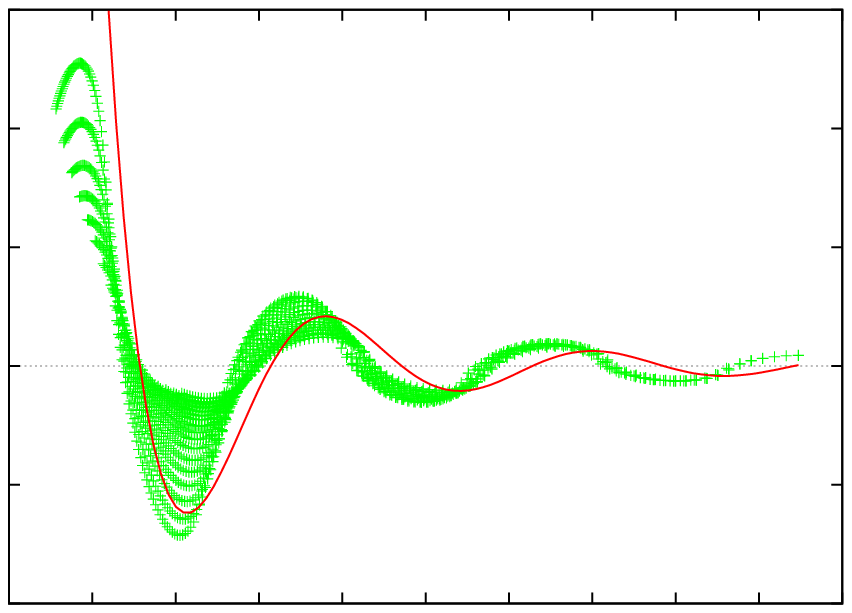}
  \caption{(colour online) \acs{IEC} energy, $E_c$, as a function of $2 \bar{k}_F d_{n + 1}$ for a superlattice with magnetic layers with a Mn doping of 2\% and 2 monolayer thickness, and a uniform impurity concentration. The (red) curve is an estimate of the ideal \acs{RKKY} range function.}
  \label{fig:Doped_2m_2pc_RKKY}
\end{figure}

There are, however, real physical reasons for deviation from \ac{RKKY} behaviour.
The data from Fig.~\ref{fig:Doped_spacer_2}(a) are replotted in Fig.~\ref{fig:Doped_2m_2pc_RKKY} as a function of $2 \bar{k}_F d_{n + 1}$. Also plotted is the function

\begin{equation}
y = \alpha \frac{\sin(x)}{x^2} ,
\end{equation}

where $\alpha$ is a scaling factor. This function is the asymptotic limit of the pseudo one-dimensional \ac{RKKY} range function.\cite{yafet_ruderman-kittel-kasuya-yosida_1987} The strength of the interaction is expected to scale with the density of states, and in the 1D case $\alpha \sim k_F^2$.\cite{dietl_free_1997}
The different series of points on the graph correspond to the series of different non-magnetic spacer thicknesses shown in Fig.~\ref{fig:Doped_spacer_2}(a). For a given $2 \bar{k}_F d_{n + 1}$, the points with the largest magnitude are those with the greatest $k_F$; this behaviour is consistent with the expected scaling of  $\alpha$ with $k_F$. Note the fact that, in order to have improved alignment of the curves, the oscillations were plotted with the effective value of the non-magnetic spacer, $d_n$, being increased in size by one monolayer, which is denoted $d_{n + 1}$. This is necessary due to the fact that the magnetic layer in this structure is not infinitely thin, as per the ideal \ac{RKKY} case, but has a defined width.

Exploring this deviation from \ac{RKKY} behaviour further, Fig.~\ref{fig:Doped_spacer_2}(b) shows the \ac{IEC} for a superlattice system with a thicker magnetic layer, $d_m / \frac{1}{2}a_0 = 8$. All other parameters are as with (a). Examining the \ac{AFM} peak, the reduction in average carrier concentration of the minimum as the spacer thickness is increased occurs more rapidly, evidenced by the larger
derivative of the average 3D carrier density, $\bar{N}_{3D}$, with respect to the non-magnetic layer thinkness, $d_n$, of the minimum $E_c$ at low $d_n$.
By way of contrast, at large $d_n$ this is lower, that is the curve has become much more straight. This is consistent with the effects of large magnetic layers increasing the centre-to-centre distance of the magnetic layers, causing the effect of an apparently larger non-magnetic layer. However, in addition to this, increasing the magnetic layer thickness has introduced additional points of inflection, for reasons that are not immediately obvious.

It is also possible to deviate from \ac{RKKY}-type behaviour through redistribution of charge. There are two primary methods by which this is achieved, as shown in Figs.~\ref{fig:GaAs_potentials}(b) and (c). The first is that charge is confined to the magnetic layers by the magnetic exchange potential. Fig.~\ref{fig:Doped_spacer_8}(a) shows the \ac{IEC} where the Mn doping has been increased to 8\%. However, when the magnetic layer is thin, significant charge redistribution is opposed by the Coulomb potential and the \ac{RKKY} character is not significantly affected.
As the figure shows, the main effect is that the size of the \ac{IEC} is increased. 
Despite the larger spin splitting and the larger polarisation of carriers caused by the greater moment concentration, the coupling retains an \ac{RKKY} character.

When the magnetic layer is made wider the increased quantity of magnetic moments now causes additional changes in the oscillatory behaviour, beyond that of simply increasing $d_m$.
Fig.~\ref{fig:Doped_spacer_8}(b) plots the \ac{IEC} for a system which now has magnetic layers of 8 monolayers with a Mn doping of 8\%.
Because of the increased depletion of carriers from the non-magnetic layers, the $N_{3D}$ values at which \ac{AFM} coupling is expected to occur are now greater for a given non-magnetic layer thickness. Additionally, the damping of the magnitude of the \ac{IEC} energy with increasing $d_n$ has now significantly changed. While the first \ac{FM} and \ac{AFM} maxima are rapidly diminished with increasing non-magnetic spacer, the second \ac{FM} peak is not greatly affected. The second \ac{AFM} peak even increases in magnitude with larger $d_n$, and for large spacer it can even be greater than the first.

Note that when the unit cell becomes large and there is a high carrier concentration, the weak coupling and flat minibands make self-consistent convergence difficult; these regions are visible as rough areas on the figures. Such samples would anyway be extremely sensitive to inhomogeneities and fluctations. No data is shown where the calculations have diverged. 

\begin{figure}
  \centering%
    \labellist%
    \footnotesize%
    \pinlabel -10 [r] at 79 126
    \pinlabel 0 [r] at 79 141
    \pinlabel 10 [r] at 79 156
    \pinlabel 20 [r] at 79 171
    \pinlabel 30 [r] at 79 186
    \pinlabel 2 [tr] at 81 121
    \pinlabel 4 [tr] at 94 108
    \pinlabel 6 [tr] at 107 95
    \pinlabel 8 [tr] at 120 82
    \pinlabel 10 [tr] at 133 69
    \pinlabel $10^{19}$ [tl] at 133 69
    \pinlabel $10^{20}$ [tl] at 178 84
    \pinlabel $10^{21}$ [tl] at 223 99
    \pinlabel {$d_n$ / $\frac{1}{2}a_0$} [tr] at 97 85
    \pinlabel {$\bar{N}_{3D}$ / cm$^{-3}$} [tl] at 193 69
    \pinlabel \rotatebox{90}{$E_c$ / \micro J\,m$^{-2}$} at 52 156
  \endlabellist%
  \subfloat[$d_m / \frac{1}{2}a_0 = 2$, 2\% Mn concentration]%
    {\includegraphics[width=0.48\textwidth]{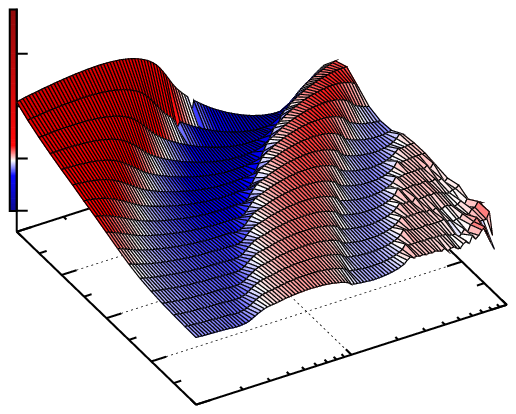}}\\
  \labellist%
    \footnotesize%
    \pinlabel -10 [r] at 79 126
    \pinlabel 0 [r] at 79 141
    \pinlabel 10 [r] at 79 156
    \pinlabel 20 [r] at 79 171
    \pinlabel 30 [r] at 79 186
    \pinlabel 2 [tr] at 81 121
    \pinlabel 4 [tr] at 94 108
    \pinlabel 6 [tr] at 107 95
    \pinlabel 8 [tr] at 120 82
    \pinlabel 10 [tr] at 133 69
    \pinlabel $10^{19}$ [tl] at 133 69
    \pinlabel $10^{20}$ [tl] at 178 84
    \pinlabel $10^{21}$ [tl] at 223 99
    \pinlabel {$d_n$ / $\frac{1}{2}a_0$} [tr] at 97 85
    \pinlabel {$\bar{N}_{3D}$ / cm$^{-3}$} [tl] at 193 69
    \pinlabel \rotatebox{90}{$E_c$ / \micro J\,m$^{-2}$} at 52 156
  \endlabellist%
  \subfloat[$d_m / \frac{1}{2}a_0 = 8$, 2\% Mn concentration]%
    {\includegraphics[width=0.48\textwidth]{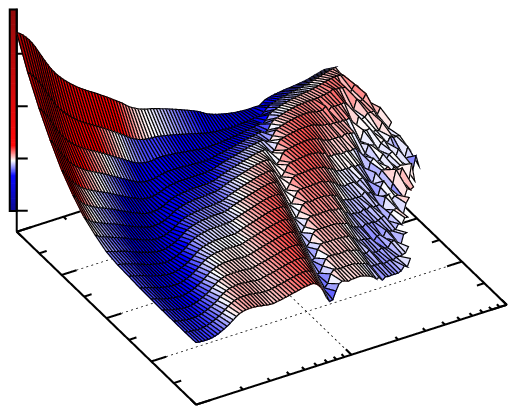}}%
  \caption{(colour online) The \acs{IEC} energy, $E_c$ of two (Ga,Mn)As/GaAs based superlattices as a function of the average 3D carrier concentration, $\bar{N}_{3D}$, and the number of monolayers of non-magnetic layer, $2 d_n / \frac{1}{2}a_0$. Both superlattices have a 2\% Mn doping in the magnetic layer but there is no charge doping in the non-magnetic spacer layer.}%
  \label{fig:Undoped_spacer}%
\end{figure}

The second method of charge redistribution is via a Coulomb potential. Fig.~\ref{fig:Undoped_spacer}(a) shows the \ac{IEC} profile for a system with a magnetic spacer of two monolayers and a Mn concentration of 2\%. However, now there is no neutralising background charge in the non-magnetic layer, so self-consistent redistribution results in the formation of an effective barrier. Fig.~\ref{fig:GaAs_potentials}(c) shows the potentials and charge distribution for a unit cell of this structure in an \ac{AFM} configuration, again with $d_n / \frac{1}{2}a_0 = 5$ and $\bar{N}_{3D} = 10^{20}$ cm$^{-3}$. As previously noted, the Coulomb barrier formed is comparable in size to the spin splitting caused by the 2\% Mn doping. This results in a similar charge redistribution as in the 8\% doped case, although without such strong carrier polarization. As with that case, there is not a significant deviation from \ac{RKKY}-type behaviour.

Increasing the magnetic spacer thickness now causes more significant changes than seen with the doped spacers. Fig.~\ref{fig:Undoped_spacer}(b) shows the \ac{IEC} profile for a superlattice with $d_m / \frac{1}{2}a_0 = 8$ with a 2\% Mn doping and no impurities in the non-magnetic spacer. In addition to the extra inflection points there is now an additional \ac{AFM} region. The magnitude of the local minimum in this region does not decrease much with non-magnetic spacer width, and occurs with an almost linear
$d\bar{N}_{3D} / d d_n$
. This is now very unlike \ac{RKKY} behaviour.

To investigate this further we shall now consider superlattice with (Al,Ga)As non-magnetic spacers, so that greater charge redistribution will occur than that caused by the magnetic ordering potential of a high magnetic moment concentration, or the Coulomb potential arising from an undoped spacer.

\subsection{(Al,Ga)As spacer}

\begin{figure}
  \centering%
  \labellist%
    \footnotesize%
    \pinlabel -10 [r] at 79 126
    \pinlabel 0 [r] at 79 141
    \pinlabel 10 [r] at 79 156
    \pinlabel 20 [r] at 79 171
    \pinlabel 30 [r] at 79 186
    \pinlabel 2 [tr] at 81 121
    \pinlabel 4 [tr] at 94 108
    \pinlabel 6 [tr] at 107 95
    \pinlabel 8 [tr] at 120 82
    \pinlabel 10 [tr] at 133 69
    \pinlabel $10^{19}$ [tl] at 133 69
    \pinlabel $10^{20}$ [tl] at 178 84
    \pinlabel $10^{21}$ [tl] at 223 99
    \pinlabel {$d_n$ / $\frac{1}{2}a_0$} [tr] at 97 85
    \pinlabel {$\bar{N}_{3D}$ / cm$^{-3}$} [tl] at 193 69
    \pinlabel \rotatebox{90}{$E_c$ / \micro J\,m$^{-2}$} at 52 156
  \endlabellist%
  \subfloat[$d_m / \frac{1}{2}a_0 = 2$, 2\% Mn concentration]%
    {\includegraphics[width=0.48\textwidth]{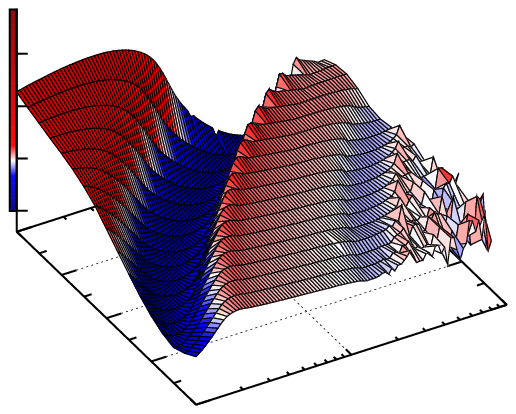}}\\
  \labellist%
    \footnotesize%
    \pinlabel -10 [r] at 79 126
    \pinlabel 0 [r] at 79 141
    \pinlabel 10 [r] at 79 156
    \pinlabel 20 [r] at 79 171
    \pinlabel 30 [r] at 79 186
    \pinlabel 2 [tr] at 81 121
    \pinlabel 4 [tr] at 94 108
    \pinlabel 6 [tr] at 107 95
    \pinlabel 8 [tr] at 120 82
    \pinlabel 10 [tr] at 133 69
    \pinlabel $10^{19}$ [tl] at 133 69
    \pinlabel $10^{20}$ [tl] at 178 84
    \pinlabel $10^{21}$ [tl] at 223 99 
    \pinlabel {$d_n$ / $\frac{1}{2}a_0$} [tr] at 97 85
    \pinlabel {$\bar{N}_{3D}$ / cm$^{-3}$} [tl] at 193 69
    \pinlabel \rotatebox{90}{$E_c$ / \micro J\,m$^{-2}$} at 52 156
  \endlabellist%
  \subfloat[$d_m / \frac{1}{2}a_0 = 8$, 2\% Mn concentration]%
    {\includegraphics[width=0.48\textwidth]{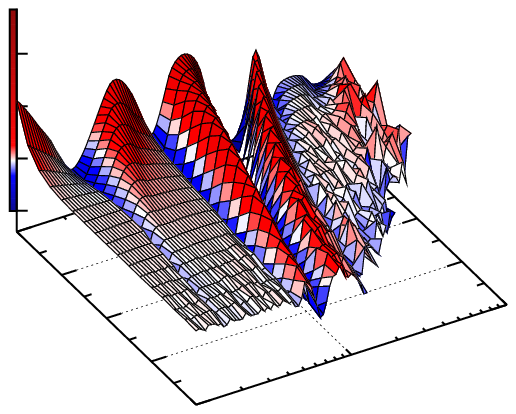}}%
  \caption{(colour online) The \acs{IEC}, $E_c$, of two (Ga,Mn)As/(Al,Ga)As based superlattices as a function of the average 3D carrier concentration, $\bar{N}_{3D}$, and the number of monolayers of non-magnetic layer, $2 d_n / \frac{1}{2}a_0$. Both superlattices have a 2\% Mn doping in the magnetic layer and the non-magnetic layers are (Al$_{0.3}$,Ga$_{0.7}$)As and have no charge doping.}%
  \label{fig:AlGaAs_spacer_2}%
\end{figure}

\begin{figure}
  \centering%
  \labellist%
    \footnotesize%
    \pinlabel -100 [r] at 79 133
    \pinlabel 0 [r] at 79 146
    \pinlabel 100 [r] at 79 159
    \pinlabel 200 [r] at 79 172
    \pinlabel 300 [r] at 79 185
    \pinlabel 2 [tr] at 81 121
    \pinlabel 4 [tr] at 94 108
    \pinlabel 6 [tr] at 107 95
    \pinlabel 8 [tr] at 120 82
    \pinlabel 10 [tr] at 133 69
    \pinlabel $10^{19}$ [tl] at 133 69
    \pinlabel $10^{20}$ [tl] at 178 84
    \pinlabel $10^{21}$ [tl] at 223 99 
    \pinlabel {$d_n$ / $\frac{1}{2}a_0$} [tr] at 97 85
    \pinlabel {$\bar{N}_{3D}$ / cm$^{-3}$} [tl] at 193 69
    \pinlabel \rotatebox{90}{$E_c$ / \micro J\,m$^{-2}$} at 52 156
  \endlabellist%
  \subfloat[$d_m / \frac{1}{2}a_0 = 2$, 8\% Mn concentration]%
    {\includegraphics[width=0.48\textwidth]{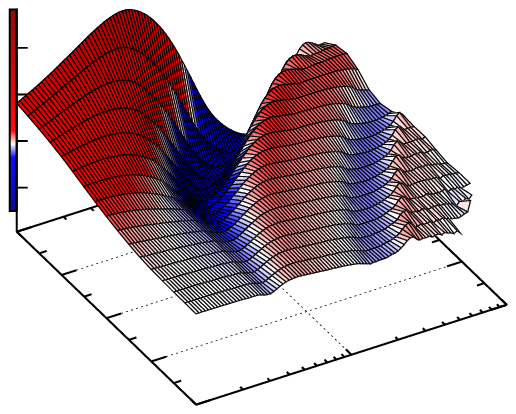}}\\
  \labellist%
    \footnotesize%
    \pinlabel -100 [r] at 79 133
    \pinlabel 0 [r] at 79 146
    \pinlabel 100 [r] at 79 159
    \pinlabel 200 [r] at 79 172
    \pinlabel 300 [r] at 79 185
    \pinlabel 2 [tr] at 81 121
    \pinlabel 4 [tr] at 94 108
    \pinlabel 6 [tr] at 107 95
    \pinlabel 8 [tr] at 120 82
    \pinlabel 10 [tr] at 133 69
    \pinlabel $10^{19}$ [tl] at 133 69
    \pinlabel $10^{20}$ [tl] at 178 84
    \pinlabel $10^{21}$ [tl] at 223 99 
    \pinlabel {$d_n$ / $\frac{1}{2}a_0$} [tr] at 97 85
    \pinlabel {$\bar{N}_{3D}$ / cm$^{-3}$} [tl] at 193 69
    \pinlabel \rotatebox{90}{$E_c$ / \micro J\,m$^{-2}$} at 52 156
  \endlabellist%
  \subfloat[$d_m / \frac{1}{2}a_0 = 8$, 8\% Mn concentration]%
    {\includegraphics[width=0.48\textwidth]{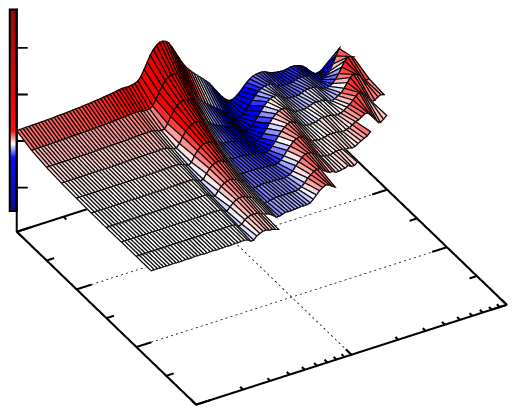}}%
  \caption{(colour online) The \acs{IEC}, $E_c$, of two (Ga,Mn)As/(Al,Ga)As based superlattices as a function of the average 3D carrier concentration, $\bar{N}_{3D}$, and the number of monolayers of non-magnetic layer, $2 d_n / \frac{1}{2}a_0$. Both superlattices have an 8\% Mn doping in the magnetic layer and the non-magnetic layers are (Al$_{0.3}$,Ga$_{0.7}$)As and have no charge doping.}%
  \label{fig:AlGaAs_spacer_8}%
\end{figure}

In the previous section it was demonstrated that interlayer coupling in superlattice structures would have an oscillatory behaviour as a function of parameters $\bar{N}_{3D}$ and $d_n$, analogous to that of \ac{RKKY}, when the magnetic layers were thin and surrounded by charge. As the structure of the superlattice is changed the \ac{IEC} would start to deviate from the ideal \ac{RKKY} behaviour. This is particularly apparent with increased magnetic layer thickness. Changing the 3D charge distribution has a more limited effect; neither large magnetic moment concentration nor a self-consistent Coulomb barrier would cause significant confinement of carriers. In order to investigate these effects further, a band offset will be introduced to further confine carriers to the magnetic layers. This will be achieved by using (Al$_{0.3}$,Ga$_{0.7}$)As as the non-magnetic layer material, which has a valence band offset of about 150 meV from GaAs.\cite{batey_energy_1986,vurgaftman_band_2001}

Fig.~\ref{fig:AlGaAs_spacer_2}(a) shows the \ac{IEC} profile for a structure with a (Ga$_{0.98}$,Mn$_{0.02}$)As magnetic layer of 2 monolayers and an (Al$_{0.3}$,Ga$_{0.7}$)As non-magnetic layer. There is no doping in the non-magnetic layers. The peak \ac{FM} and \ac{AFM} coupling strengths are now stronger than in the case with doped GaAs spacers seen in the otherwise identical structure in Fig.~\ref{fig:Doped_spacer_2}(a).
The charge distribution of this structure is shown in Fig.~\ref{fig:AlGaAs_potentials}(a), where the barrier confines carriers to the magnetic layers, as expected. However, the $2 d_n \bar{k}_F$ oscillations are damped more rapidly than with the GaAs spacer, resulting in the second \ac{FM} and \ac{AFM} peaks being very weak. This additional damping occurs particularly rapidly with increasing carrier density, $\bar{N}_{3D}$. As a result, the first anti-ferromagnetic peak barely reduces in magnitude as the non-magnetic layer thickness is increased.
Juxtaposing this with the GaAs barrier case, where the largest \ac{AFM} coupling that can occur when $d_n / \frac{1}{2}a_0 = 10$ is less than a quarter of the size of that when $d_n / \frac{1}{2}a_0 = 2$, we identify this as a significant departure from the previously seen \ac{RKKY}-like oscillatory behaviour.

Increasing the magnetic moment concentration leads to a more interesting alteration than with the GaAs spacer, where the effect was principally to scale up the magnitude of the \ac{IEC} energy. Fig.~\ref{fig:AlGaAs_spacer_8}(a) shows the \ac{IEC} for a (Ga,Mn)As/(Al,Ga)As superlattice with an 8\% Mn doping in the 2 monolayer magnetic layer, as previously considered in Fig.~\ref{fig:AlGaAs_potentials}(b). Now the first \ac{AFM} peak appears to have two stages. The first is at low spacer thicknesses, where the average hole density at which the maximum occurs decreases with increasing spacer thickness. For large spacer thicknesses the curve has straightened out, and there is almost no dependence on $d_n$ for the sign of the coupling. This characteristic is similar to that exhibited in Figs.~\ref{fig:Doped_spacer_2}(b) and~\ref{fig:Doped_spacer_8}(b), where the magnetic layer is 8 monolayers thick. This was attributed to loss of independence of the $d_n$ and $\bar{N}_{3D}$ parameters, as the system became less \ac{RKKY}-like. 
Knowing that the band offset and large magnetic ordering cause significant carrier redistribution, particularly, this means that the carrier concentration in the spacer will decrease as a function of spacer thickness. This can account for the weak dependence of $E_c$ on $k_F d_n$. Also, note that the size of the first \ac{AFM} peak decreases more rapidly at high spacer thicknesses where the average carrier concentration, $\bar{N}_{3D}$, at which it occurs is not decreasing. This is consistent with the previous observation of enhanced damping with increasing carrier concentration.

With high magnetic layer thicknesses the \ac{RKKY}-type oscillations have almost completely disappeared. The beating patterns which were emerging in the $d_m / \frac{1}{2}a_0 = 8$ GaAs spacer cases have now come to dominate the \ac{IEC}. Figs.~\ref{fig:AlGaAs_spacer_2}(b) and~\ref{fig:AlGaAs_spacer_8}(b) shows this for $d_m / \frac{1}{2}a_0 = 8$, with respectively 2\% and 8\% Mn doping. In these cases the oscillations occur almost exclusively as a function of hole density, being almost independent of the spacer thickness. Note, however, as was seen in  Fig.~\ref{fig:AlGaAs_potentials}(c), the non-magnetic layer is highly depleted when the magnetic layer is 8 monolayers thick with an 8\% Mn doping. This makes computing \ac{IEC} for larger spacers unfeasible.

\section{Discussion and recipes}

Having explored the parameter spaces we will now consider possible structures of a (Ga,Mn)As based superlattice that would exhibit \ac{AFM} interlayer coupling. Each parameter will be considered for feasibility, and, based on the above calculations, suggestions for values can be made.

The first to be considered is the Mn concentration in the (Ga,Mn)As layers. From the viewpoint of simply creating a viable ferromagnet this is an essential parameter; not only does each substitutional Mn provide a magnetic moment, it also acts as an acceptor and thus this factor controls the hole concentration. Calculations\cite{jungwirth_prospects_2005} estimate that the minimum hole density for ferromagnetism is $\sim 10^{20}$ cm$^{-3}$. Assuming that each Mn provides one hole, this carrier concentration would correspond to a moment concentration of $\sim 0.5\%$. Experimentally, typical Mn concentrations are in the range of 2-8\% ($4.4 \times 10^{20}$ to $1.8 \times 10^{21}$ cm$^{-3}$), which in good quality material could result in higher carrier concentrations than the $10^{19}$ to $10^{21}$ cm$^{-3}$ range considered in these calculations. Whilst the higher magnetic moment concentration can increase the size of the $E_c$ peak, and thus a high moment concentration is favourable, the high carrier concentrations that would be associated with this would cause the strength of the \ac{IEC} to become extremely weak. This constraint therefore imposes a practical range for Mn concentrations as being between 2 and 4\% ($4.4 \times 10^{20}$ to $8.8 \times 10^{20}$ cm$^{-3}$ respectively).

For the non-magnetic spacer thickness the general trend is that the strength of the \ac{IEC} becomes weaker as the non-magnetic layer becomes thicker. Although this effect is somewhat diminished for the cases where there is strong carrier confinement to the magnetic layers, it is a serious consideration and, ideally, to see strong \ac{IEC} effects, the non-magnetic layer should be as thin as possible. Furthermore, particularly in cases where the $2 k_F d$ behaviour is dominant, as carrier concentration increases the spacer thickness at which the \ac{AFM} \ac{IEC} is strongest decreases inversely. As discussed above, low carrier concentrations are not possible, so therefore it would seem beneficial to make the spacer layers as thin as practicable. Bearing in mind that the average distance between two Mn atoms when the concentration is 3\% is of the order of a couple of GaAs unit cells, in order to make the non-magnetic spacer a discernible barrier then 4 monolayers would seem to be a realistic lower bound.

The effect of the magnetic layer thickness on the \ac{IEC} profile is more subtle, and seems mainly to distort the \ac{RKKY} behaviour but otherwise, in the limits considered within this study, does not have any negative effects on the interlayer coupling. However, again for interlayer coupling to exist it is necessary that each magnetic layer is itself ferromagnetic. Usually (Ga,Mn)As is grown in bulk layers of many nanometres; the thinnest (Ga,Mn)As epilayers for which published literature exists are 5 nm thick,\cite{giddings_large_2005} and the (Ga,Mn)As based heterostructures with magnetic layers as thin as 8 monolayers have been reported to be ferromagnetic.\cite{sadowski_ferromagnetic_2002} It would therefore seem prudent, in order to ensure that the magnetic layers are effective ferromagnets, to prefer to make them thicker. For the 5 nm film some amount of surface depletion should be expected, so a 8 monolayer thick magnetic layer, equivalent to 2.26 nm, is comparable. Of course, if thinner films are shown to be viable then there is no reason not to consider them also.

\begin{figure}
  \centering
  \labellist%
    \footnotesize%
    \pinlabel $10^{19}$ [B] at 81, 60
    \pinlabel $10^{20}$ [B] at 201, 60
    \pinlabel $10^{21}$ [B] at 321, 60
    \pinlabel {$\bar{N}_{3D}$ / cm$^{-3}$} [B] at 201, 47
    \pinlabel $-20$ [r] at 81, 73
    \pinlabel $-15$ [r] at 81, 94
    \pinlabel $-10$ [r] at 81, 115
    \pinlabel $-5$ [r] at 81, 136
    \pinlabel $0$ [r] at 81, 157
    \pinlabel $5$ [r] at 81, 178
    \pinlabel $10$ [r] at 81, 199
    \pinlabel $15$ [r] at 81, 220
    \pinlabel $20$ [r] at 81, 241
    \pinlabel \rotatebox{90}{$E_c$ / \micro J\,m$^{-2}$} at 56, 157
    \hair 4pt
    \pinlabel GaAs [r] at 165, 232
    \pinlabel (Al$_{0.3}$,Ga$_{0.7}$)As [r] at 165, 220
  \endlabellist%
  \includegraphics[width=0.48\textwidth]{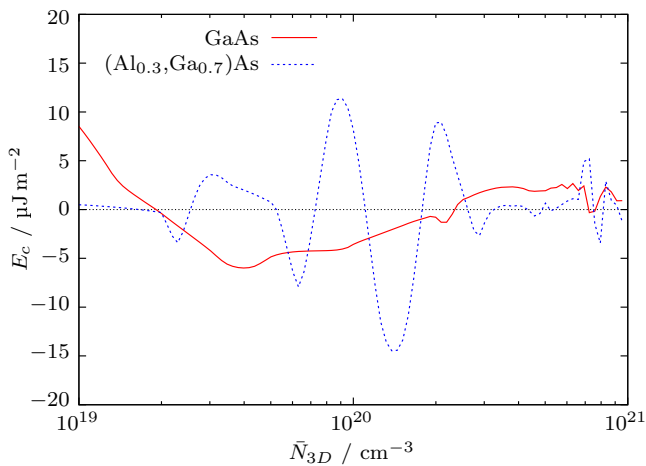}
  \caption{(colour online) A comparison of the \acs{IEC} energy, $E_c$, as a function of the average 3D carrier concentration, $\bar{N}_{3D}$, for two specific superlattices with either a GaAs or an (Al$_{0.3}$,Ga$_{0.7}$)As non-magnetic layer. The magnetic layers are 8 monolayers thick and have a Mn concentration of $5 \times 10^{20}$ cm$^{-3}$ (2.26\%) and the non-magnetic layer is 4 monolayers thick.}
  \label{fig:grown_layers_reduced}
\end{figure}

Based on these constraints, Fig.~\ref{fig:grown_layers_reduced} shows the \ac{IEC} profile for two candidate superlattices as a function of carrier concentration, $\bar{N}_{3D}$. Both superlattices are identical in structure except for the composition of the non-magnetic layer. The magnetic layer thickness is 8 monolayers and has a magnetic impurity concentration of $5 \times 10^{20}$ cm$^{-3}$ (2.26\%) and the non-magnetic layers are 4 monolayers thick. As expected from the calculations, when the (Al,Ga)As barriers strongly confine carriers to the magnetic layers the \ac{IEC} energy can have potentially greater magnitudes, although the oscillations have a much higher frequency. In these samples the carrier concentration would be somewhere below the Mn concentration of $5 \times 10^{20}$ cm$^{-3}$ (2.26\%), however the exact amount would depend on subtleties of the growth conditions. Although this suggests that for \ac{AFM} \ac{IEC} to occur the desired carrier concentration should be several times lower, it must be accepted that the calculations are of a more qualitative nature. Additionally, by tailoring the band offset of the non-magnetic layer by altering the Al content, the location of the peak can be adjusted somewhat.
This at least shows that these designs offer the possibility for \ac{AFM} interlayer coupling.

Even if the \ac{IEC} energy were to favour an \ac{AFM} arrangement, if the \ac{AFM} coupling is weaker than the anisotropy fields it is possible that, after the application of a field, the superlattice could become locked into a \ac{FM} spin configuration. This spin-locking behaviour has been observed in EuS/PbS superlattices\cite{kepa_antiferromagnetic_2001} and Fe/Nb multilayers\cite{rehm_magnetic_1997} studied via neutron scattering. 

Comparing, then, the calculated \ac{IEC} to the magnetocrystalline anisotropic energy of (Ga,Mn)As, we take a typical ``worst case'' value of the in-plane cubic anisotropy constant to be of the order of 2000 J m$^{-3}$ at 4.2 K.\cite{wang_spin_2005}
Using a value of the interlayer coupling energy $E_c = 10~\micro$J$\,$m$^{-2}$ from Fig.~\ref{fig:grown_layers_reduced} and using the bilayer period of 3.4 nm we find the energy density of the \ac{IEC} energy is 3000 J m$^{-3}$. Although this is assuming an ideal value of $E_c$, this compares favourably with the anisotropy energy. Furthermore, larger values for the \ac{IEC} have been found in the tight-binding approach.\cite{sankowski_interlayer_2005}
Therefore, such a superlattice structure might reasonably be expected to be a candidate to exhibit \ac{AFM} interlayer coupling.

\section{Conclusion}

The composition and structure of (Ga,Mn)As based superlattices can have profound effects on the expected \ac{IEC}. By examining possible compositions within the broad parameter space that these structures offer it is possible to identify different recipes for devices that might offer the possibility of demonstrating \ac{AFM} interlayer coupling. Such a study was presented in this paper based on the parabolic band ${\bf k} \cdot {\bf p}$ kinetic exchange model. This model ignores spin-orbit and band warping effects, although comparisons to the previously studied microscopic tight-binding model suggest that our results provide a reasonable qualitative or semi-quantitative description of the system. The calculations predict that short period superlattices with magnetic and non-magnetic layers with widths less than 10 monolayers seem be promising candidates. There is existing experimental work in (Ga,Mn)As/GaAs based superlattices with similar dimensions, but this has only exhibited \ac{FM} \ac{IEC}.\cite{szuszkiewicz_interlayer_1998,sadowski_ferromagnetic_2002} The ideal dimensions suggested by the calculations are therefore feasible and have shown themselves to be viable ferromagnetic \ac{DMS} material.

Particularly interestingly, in a (Ga,Mn)As/(In,Ga)As based superlattice composed of 8 and 4 monolayer width layers respectively two phase transitions were observed.\citep{hernandez_magnetic_2001} Although there was no evidence that this was due to any \ac{AFM} effects, this indicates that there is some additional physics at play in these systems, making them of interest for further study. The calculations presented in this paper have shown the importance of the composition of the non-magnetic spacer on the character of the \ac{IEC}. Using different Al concentrations to tailor the band-offset between layers in a superlattice is a standard tool for designing normal non-magnetic superlattice systems. Utilising this technique in the magnetic superlattices could potentially provide a way to tune to \ac{IEC} profile to one where \ac{AFM} coupling is preferential.

We acknowledge support from EU Grant IST-015728, from UK Grant GR/S81407/01, from CR Grants KAN400100652, FON/06/E002, AV0Z1010052, and LC510.

\acrodef{AFM}{anti\acl{FM}}
\acrodef{DMS}{dilute magnetic semiconductor}
\acrodef{FM}{ferromagnetic}
\acrodef{GMR}{giant \acl{MR}}
\acrodef{IEC}{interlayer exchange coupling}
\acrodef{LSDA}{local-spin density approximation}
\acrodef{MR}{magnetoresistance}
\acrodef{RKKY}{Ruderman-Kittel-Kasuya-Yosida}


\end{document}